\documentclass[12pt,draftclsnofoot,onecolumn]{IEEEtran}

\usepackage[tbtags]{amsmath} 
\usepackage{amssymb}  
\usepackage{verbatim} 
\usepackage{amsxtra}  

\usepackage{enumerate}

\usepackage{amsfonts}
\usepackage{color}

\usepackage{cite}

\usepackage{mathtools}
\mathtoolsset{showonlyrefs=true}

\usepackage{url}

\usepackage{epsfig, graphicx, psfrag}

\usepackage[mathcal]{euscript}
\newenvironment{frcseries}{\fontfamily{frc}\selectfont}{}

\usepackage{ifthen}
\providecommand{\VersionLength}{long}
\newcommand{\ver}{\ifthenelse{\equal{\VersionLength}{long}}}
\newcommand{\nver}{\ifthenelse{\equal{\VersionLength}{short}}}
\newcommand{\col}{\ifthenelse{\equal{\ColumnNum}{1}}}

\providecommand{\second}[2]{#2}

\newtheorem{remark}{Remark}

\providecommand{\secref}[1]{Section~\ref{#1}}

\providecommand{\figref}[1]{Figure~\ref{#1}}

\newcommand{\ie}{i.e.}
\newcommand{\eg}{e.g.}

\newcommand{\viz}{viz.}
\newcommand{\etal}{et al.}
\newcommand{\cf}{cf.}

\newcommand{\bm}[1]{\mbox{\boldmath{$#1$}}}

\newcommand{\SNR}{\text{SNR}}

\newcommand{\mI}{\mathcal{I}}

\newcommand{\Comment}[1]{}
\newcommand{\old}[1]{}
\newcommand{\rem}[1]{}

\newcommand{\by}{\bm y}

\newcommand{\bH}{\text{\bf H}}
\newcommand{\bI}{\text{\bf I}}

\newcommand{\bK}{\text{\bf K}}

\newcommand{\bx}{{\bm x}}

\newcommand{\bz}{{\bm z}}

\providecommand{\comment}[1]{}

\newcommand{\MAC}{{\text{MAC}}}

\newcommand{\beqn}[1]{\begin{eqnarray}\label{#1}}
\newcommand{\eeqn}{\end{eqnarray}}
\newcommand{\beq}[1]{\begin{equation}\label{#1}}
\newcommand{\eeq}{\end{equation}}

\newcommand{\cC}{{\mathcal C}}

\makeatletter
\newcommand{\vast}{\bBigg@{4}}
\newcommand{\Vast}{\bBigg@{5}}
\makeatother

\providecommand{\C}[1]{\text{C} \left( #1  \right)}

\providecommand{\inter}{{\text{inter}}}
\providecommand{\bHinter}{\bH_{\text{inter}}}
\providecommand{\bxinter}{\bx_{\text{inter}}}

\providecommand{\Cmac}[1]{ \cC_\MAC \left( #1 \right) } 

\providecommand{\dB}{\mathrm{dB}}
\providecommand{\self}{\textrm{self}}
\providecommand{\intra}{\textrm{intra}}
\providecommand{\inter}{\textrm{inter}}

\usepackage{hyperref}

\providecommand{\ColumnNum}{1}

\begin{document}

\title{Layered Uplink Transmission in \\ Clustered Cellular Networks}

\author{Anatoly Khina, Tal Philosof and Moshe Laifenfeld}

\maketitle


\begin{abstract}
    The demand for higher data rates and the scarce spectrum resources drive the adoption of collaborative communication techniques. 
    In this work we shown that the existing cluster based collaborative schemes can be greatly improved in terms of both the achievable performance, and complexity, 
    by allowing overlapping across clusters. Further improvement is achieved by incorporating scheduling to the decoding process. 
    Different variants of the improved schemes are considered and are shown to achieve near optimum performance for practical signal-to-noise ratios for the one- and two-dimensional (hexagonal) Wyner-type models.
\end{abstract}

\allowdisplaybreaks

\section{introduction}
\label{s:intro}

The growing demand for high data rates on the one hand and the lack of spectrum resources on the other hand leads many cellular networks to follow two main trends: (I) Operation over a single frequency band across the entire network~--- known as frequency reuse one; (II) Moving to heterogeneous networks, which, in addition to the legacy macro-cells incorporate small cells, \eg, micro-, pico- and femto-cells~\cite{HetNet}.
Small cells increase the capacity of the network by increasing spatial reuse; however, this gain is limited by interferences which are inevitable in frequency reuse one paradigm. The mutual interference is more pronounced at the cell edge, where mobile users that are connected to different base stations interfere with each other on both the uplink and downlink directions.

To reduce these mutual interferences, several schemes were adopted by the standardization bodies that are based on a cooperation between neighboring base stations. The Inter-Cell Interference Coordination (ICIC) techniques \cite{Survey_ICIC} reduces mutual interferences by coordinating neighboring base-stations resources allocation to cell-edge users. Coordinated Multi-Point(CoMP) reception and transmission \cite{COMP_3GPP} is another technique not only aiming at reducing interferences, but also increasing data-rates through multi-site joint transmission, coordinated beamforming and coordinated scheduling among users.
Cooperation among base stations is expected to continuously grow, with the introduction of the future cellular generations (the \emph{fifth generation}, 5G), along with the costs it incur. The increasingly high computational complexity that scales exponentially with the number of users, and which is expected to grow even further with the adoption of sophisticated multiuser MIMO techniques, in combination with the increasing demands on throughput, low latency, and high reliability on the backhaul links interconnecting base-stations (for all cell sizes), drive infrastructure costs to skyrocket.

All these collaboration techniques require high throughput, low latency, and high reliability backhaul links between base-stations for all cell sizes, which pose a challenge to the network architects, both in terms of complexity and cost. To reduce the increasing infrastructure costs  a \emph{cloud based architecture} \cite{Cloud_China_Mobile} is considered by the industry and standardization bodies. A cloud based cellular architecture is based on an already gaining traction concept of breaking the classical co-located radio frequency (RF) and the base-band processing, into (multiple) Remote Radio Heads (RRHs) distributed in the cell's coverage area, and
a single Base-Band Unit (BBU), connected via high throughput, low latency \emph{fronthaul} links, carrying, traditionally, in-phase and quadrature samples.
The cloud architecture takes this concept one step further, introducing the Radio Access Network (RAN) or the \emph{Cloud RAN}, by pushing the BBUs to the cloud to form a highly computationally capable central unit. Interference mitigation schemes, as well as rate enhancement schemes can now achieve better performance as they are designed in a centralized manner \cite{CloudRAN_2014_let,Poor14_CRAN,Layered_CloudRAN}.
Specifically, Information theoretical aspects for compression of frounthaul links in cloud RAN architecture are considered in \cite{CloudRAN_2014}.

\begin{figure}
\col{
\centering
	\includegraphics[width=.7\columnwidth]{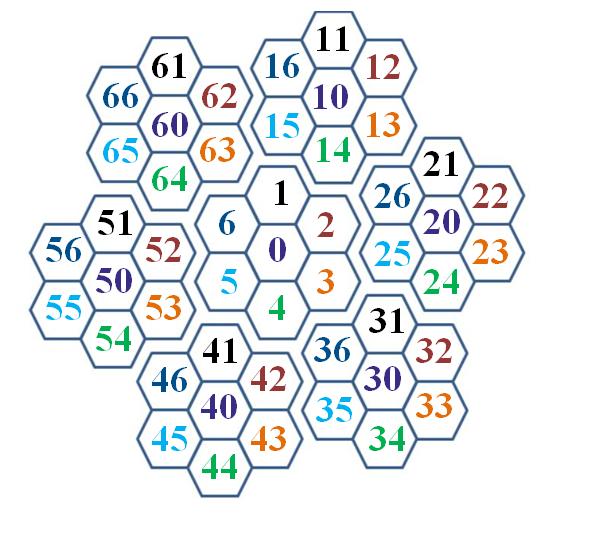}
}{
\centering
	\includegraphics[width=\columnwidth]{./figures/cell2d.jpg}
} 
	\caption{Two-dimensional clustered cell-array \cite{KatzZaidelShamai:ISIT2013}. Each cluster is composed of a central cell and six surrounding ``peripheral'' cells.}
\label{fig:HK:2D}
\end{figure}

In this study, we focus on collaborative schemes for enhancing the uplink rates in a cellular network.
In the sequel, we analyze several such schemes, which require different levels of collaboration between base-stations. Our analysis builds upon well established models and theoretic results from the literature. Due to brevity, we mention here only a few.

Wyner's seminal work \cite{Wyner94} is a pioneer study of BS collaboration for uplink communication, for one- and two-dimensional (hexagonal array) cellular.
The model in \cite{Wyner94} assumes one user per BS, where each BS receives the uplink signal of its own cell along with those of its immediate neighbors attenuated by some constant factor. He further derived the uplink capacity through joint decoding of all cell signals with no limitation on the backhaul capacity.
This result can therefore be used as a theoretical upper bound for any collaborative or cloud RAN scheme. In practice, however, there is a trade-off between backhaul/fronthaul bandwidth, computational complexity and the uplink achievable data rates.

Katz \etal~\cite{KatzZaidelShamai:ISIT2013} proposed a collaborative scheme for the two-dimensional model, that partitions the entire cellular network into fixed disjoint clusters,
consisting of a central cell surrounded by its six neighbors, as depicted in \figref{fig:HK:2D}; 
within each cluster, the central BS is connected to the peripheral cells via six backhaul links.
Joint decoding of all cell messages is carried by the central cell. 
The basic building block used in \cite{KatzZaidelShamai:ISIT2013} is the renowned layered transmission scheme by Han and Kobayashi \cite{HanKobayashi} over the interference channel.
Han and Kobayashi demonstrated that partial decoding of the interference signal can boost the rate of the desired message.
Similarly, Katz \etal~\cite{KatzZaidelShamai:ISIT2013} used layered transmission to enhance performance by partially decoding interference from neighboring clusters. 
Clearly, the central cell within a cluster enjoys a better performance than the rest due to a more complete information available from its immediate neighboring cells.

In this work, we introduce two new techniques: \emph{cluster overlap} and \emph{scheduled decoding}. 
The cluster overlap scheme enhances the clustered scheme of \cite{KatzZaidelShamai:ISIT2013} by essentially allowing each cell to act as the central cell in their work, 
which means, in turn, that all cells enjoy the same performance.
Further enhancement is attained by noting that, in contrast to the non-overlap clustering scheme where the messages of the peripheral cells had to be recovered completely, 
in the cluster overlap scheme, only the message of the central cell needs to be recovered; 
this difference promises great boost in performance for small interference attenuation coefficients in Wyner's model even for simplified variants of the scheme.
The downside of this approach is the increase in fronthaul/backhaul communication, as six links are required per cell, instead of the single link per cluster in the non-overlap counterpart. 

Our second proposed technique is that of scheduled decoding, namely, iterative decoding of the cell messages according to a predefined order.
Specifically, the network cells are partitioned into finite disjoint sets of cells (usually small number of sets),
where in each \emph{decoding iteration} all the cells in a specific set are decoded simultaneously. 
The recovered messages of these cells are then communicated over the backhaul/fronthaul links to their neighbors, 
who can subsequently decode their message after subtracting the interference coming from cells decoded at a previous phase of the schedule.
The decoding process is completed after decoding all sets. This technique can be implemented either over cloud RAN, 
where a trade-off between performance and computation complexity is of the essence, 
or as a collaborative network solution based on backhaul/fronthaul links between BSs, trading off performance and backhaul bandwidth by sharing either soft decisions among sets (clustering) or hard decisions (scheduling) to reduce backhaul throughput.
We demonstrate that the rate achieved by this scheme can give substantial performance boosts, and, in conjunction with the cluster overlap technique achieves close-to-capacity performance (Wyner's upper bound) for the parameter used and encountered min practice.

The rest of the paper is organized as follows. In the next section, we introduce the system model for one dimensional cellular network and two dimensional hexagonal cells.
The results by Wyner~\cite{Wyner94} for full cooperation with unlimited backhaul capacity between base stations is presented \secref{s:Wyner}.
In \secref{s:1cell}, we derive achievable rates when no backhaul is available.
\secref{s:HK} is a recap on non-overlap clustering approach \cite{KatzZaidelShamai:ISIT2013}.
Clustering with overlaps is introduced in \secref{s:overlap}.
The simplest scheduled decoding variant~--- time-sharing~--- is presented in \secref{s:time-sharing}.
Other scheduled decoding techniques that require only \emph{digital backhaul} are discussed in \secref{s:tetris:1antenna}.
A scheme that uses both clustering (analog backhaul) and schedueld decoding (digital backhaul) is considered in \secref{s:tetris}.
Further imporvements via multi-round scheduling are discussed in \secref{s:moses}.
A performance evaluation and comparison are shown in \secref{s:numeric}\second{, 
followed by a discussion in \secref{s:discuss}}{}.


\section{System Model}
\label{s:model}

In this section we present the one- and two-dimensional cellular network models as
proposed by Wyner~\cite{Wyner94}. 


\begin{figure}
    \centering
	\includegraphics[width=.8\columnwidth]{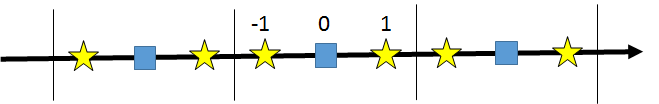}
	\caption{One-dimensional clustered cell-array. Each cluster is composed of a central cell and two surrounding ``peripheral'' cells.}
\label{fig:HK:1D}
\end{figure}

In the one-dimensional model, depicted in \figref{fig:HK:1D},
each cell (indexed by $0$) has two neighboring cells (indexed $\pm1$), which cast an interference multiplied by a channel gain $\alpha$: 
\begin{align}
\label{eq:channel:1D}
    y_0 = x_0 + \alpha x_{-1} + \alpha x_1 + \bz_0
    ,
\end{align}
where $x_0, x_1,$ and $x_{-1}$ are the signals of cells $0, 1$  and $(-1)$ of average power $P$,\footnote{Throughout this paper, we treat all messages coming from different users within a specific cell as a single message signal.}
and $\bz_0$ is an additive white Gaussian noise (AWGN) of unit power.

The two-dimensional model is composed of cells lying on a hexagonal lattice, as depicted in \figref{fig:HK:2D}:
Each cell (indexed 0) has six neighboring cells (indexed $1,\ldots,6$):
\begin{align}
\label{eq:channel:2D}
    y_0 = x_0 + \alpha \sum_{i=1}^6 x_i + \bz_0 ,
\end{align}
where $x_i$ is the signal of cell $i$ of average power $P$, and $z_0$ is AWGN of unit power.

As implied by \eqref{eq:channel:1D} and \eqref{eq:channel:2D}, we shall make use of the notion of MISO multiple-access channel (MAC), throughout this work:
\begin{align}
	\by = \bH \bx + \bz ,
\end{align}
where $\by$ denotes the channel output vector of length $N_r$, $\bz$ is a Gaussian noise vector of covariance $\bK$,
$\bx$ is an input vector of length $N_t$, where each of its entries corresponds to an independent codebook of power $P$ transmitted by a different single-antenna transmitter,
and $\bH$ is the channel matrix of dimensions $N_r \times N_t$.
The achievable rate region of the $N_t$ codebooks/users $(R_1, \ldots, R_{N_t})$ is given (see~\cite{YuPhD}) by
\begin{subequations}
\begin{align}
	&\Cmac{ \bH, \bK } \triangleq
	\Bigg\{
		(R_1, \ldots, R_{N_t}) :
\label{eq:MAC_capacity:tuple}
	\\ &\qquad
		\sum_{i \in \mI} R_i \leq \log \frac{ \left| \bK + P \bH_\mI \bH_\mI^T \right| }{ \left| \bK \right| }
		,
		 \mI \subseteq \{1, \ldots, N_t\}
	\Bigg\} ,
\label{eq:MAC_capacity:constraints}
\end{align}
\end{subequations}
where $\bH_\mI$ denotes the sub-matrix composed of the columns of $\bH$ of indices in $\mI$.
Note that the number of users, $N_t$, is determined by the number of columns of $\bH$.


\section{Full Backhaul and Unrestricted Complexity}
\label{s:Wyner}

Optimal rate is achieved when all cells convey their (``raw'') measured signals \eqref{eq:channel:1D} to a centralized processing unit in centralized base-station or \emph{cloud} based computing.
Using these measurements, the cloud recovers the messages of all the cells.

The achievable rate (of each cell) is \cite{Wyner94}:
\begin{align}
    C_\text{Wyner} = \int_0^1 \C{ \left[ 1 + 2 \alpha \cos(\tau \theta) \right]^2 P } d\theta \,,
\end{align}
where $\tau = 2 \pi$ is the circle constant~\cite{PiIsWrong_Original,PiIsWrong_NoReally} and\footnote{All logarithms are taken to the base 2 and all rates are measured in bits.}
\begin{align}
    \C{x} &\triangleq \frac{1}{2} \log(1 + x)
\end{align}
is the capacity of a scalar AWGN channel of SNR $x$.

Unfortunately, the computational complexity, backhaul requirements and latency constraints make this setting infeasible when the cellular network size is large.
In the sequel, we consider schemes with reduced complexity and backhaul/fronthaul requirements, and compare their rate to $C_\text{Wyner}$.


\section{No Backhaul}
\label{s:1cell}

In this section we make no use of backhaul communication. Namely, each cell decodes its intended message with no cooperation with other cells.

One can devise a simple scheme by treating all interference coming from neighboring cells as Gaussian noise.\footnote{Replacing the interfering noise with a Gaussian one having the same power is justifiable both theoretically, since it is the worst possible additive noise with a given power (\cf~\cite{CoverBook2Edition}) and in practice (see~\cite{LeeBook3rd}).}
This scheme achieves the following rate
\begin{align}
\label{eq:naive_rate}
    R_\text{Na\"ive} = \C{ \frac{P}{1 + 2 \alpha^2 P} } .
\end{align}

The na\"ive approach
can be improved by using the partial decoding technique proposed by Han and Kobayashi for the interference channel~\cite{HanKobayashi}.
According to this technique, each cell partially recovers the interference coming from its neighboring cells (while treating the residual non-recovered interference as Gaussian noise) to facilitate the (full) recovery of its intended message.
To this end, split each of the cell signals $x_i$ into two parts:
\begin{align}
    x_i = x_i^d + x_i^{ud}
\end{align}
where signals $x_i^d$ ($i=-1,0,1$) are decoded at cell 0; the signals $x^{ud}_i$ for $i = \pm 1$, are treated as noise, whereas $x^{ud}_0$ is recovered at cell 0.
The power allocated to $x_i^d$ is $\lambda P$ ($0 \leq \lambda \leq 1$), whereas the remaining power $(1-\lambda) P$ is allocated to $x_i^{ud}$.

The optimal achievable rate of this scheme is equal to 
\begin{align}
	R = \max_{R^d, R^{ud}} \left\{ R^d + R^{ud} \right\},
\end{align}
where $R^d$ and $R^{ud}$ are the rates corresponding to $x_i^d$ and $x_i^{ud}$, respectively,
and the maximization is carried over all rates satisfying
\begin{align}
\label{eq:pDF:Cmac}
\begin{aligned}
	&\left( R^d, R^d, R^{ud}, R^d \right) \in
\noeqref{eq:pDF:Cmac:region}
\col{}{\\
	&\:\:\:\:} \Cmac{
		P [\alpha \sqrt{\lambda}, \sqrt{\lambda}, \sqrt{1 - \lambda}, \alpha \sqrt{\lambda}] ,
	 	1 + 2 (1 - \lambda) P
	}
.
\end{aligned}
\end{align}
The rate region of \eqref{eq:pDF:Cmac} can be explicitly written as:
\begin{subequations}
\label{eq:1D:1cell:pDF}
\begin{align}
    R^{ud} &\leq \C{ \frac{(1-\lambda)P}{ 1 + 2 \alpha^2 (1-\lambda) P } }
\label{eq:1D:1cell:pDF:Rud}
 \\ R^{d} &\leq \frac{1}{i} \C{ \frac{i \alpha^2 \lambda P}{ 1 + 2 \alpha^2 (1-\lambda) P } }, \quad i = 1, 2
\label{eq:1D:1cell:pDF:Rd_i}
 \\ R^{d} &\leq \frac{1}{3} \C{ \frac{(1 + 2 \alpha^2) \lambda P}{ 1 + 2 \alpha^2 (1-\lambda) P } }
\label{eq:1D:1cell:pDF:Rd_all}
 \\ i R^d + R^{ud} &\leq \C{ \frac{(1-\lambda) P + i \alpha^2 \lambda P}{ 1 + 2 \alpha^2 (1-\lambda) P } }, \quad i = 1, 2
\label{eq:1D:1cell:pDF:Rd_i_and_Rud}
 \\ 3 R^d + R^{ud} &\leq \C{ \frac{P + 2 \alpha^2 \lambda P}{ 1 + 2 \alpha^2 (1-\lambda) P } } \,.
\label{eq:1D:1cell:pDF:Rd_all_and_Rud}
\end{align}
\end{subequations}

By using simple geometrical considerations, one verifies that the optimal sum-rate is achieved for $R^{ud}$ that satisfies
\eqref{eq:1D:1cell:pDF:Rud} with equality. Furthermore, one can easily show that \eqref{eq:1D:1cell:pDF:Rd_i_and_Rud} is more stringent than \eqref{eq:1D:1cell:pDF:Rd_i},
and the limiting inequality is attained for $i=2$. The same holds for \eqref{eq:1D:1cell:pDF:Rd_all_and_Rud} and \eqref{eq:1D:1cell:pDF:Rd_all}.
Thus, the achievable rate using this scheme is
\begin{subequations}
\begin{align}
    R \col{}{&=} \max_{\lambda \in [0,1]} \Bigg[ \col{&}{}\C{ \frac{(1-\lambda)P}{ 1 + 2 \alpha^2 (1-\lambda) P } }
    \col{}{\\ &\ \ \ } + \min \bigg\{ \frac{1}{2} \C{ \frac{ 2 \alpha^2 \lambda P }{ 1 + (1 + 2 \alpha^2) (1-\lambda) P } },
    \\ &\col{}{\ \ \ \ \ \ \ \ \ \ \ \ \ \ } \frac{1}{3} \C{ \frac{(1 + 2 \alpha^2) \lambda P}{ 1 + (1 + 2 \alpha^2) (1-\lambda) P } } \bigg\}
    \Bigg]
    \,.
\end{align}
\end{subequations}


\section{Non-Overlapping Clustering}
\label{s:HK}

An intermediate approach between the two extremes~--- full and no backhaul~--- was proposed by Katz et al.\ \cite{KatzZaidelShamai:ISIT2013} for the two-dimensional model. 
According to this approach, the whole cell array is divided into clusters,
where each cluster is composed of seven cells: a central cell surrounded by its six immediate neighbors, which will be referred to as ``peripheral cells''.
All messages within the cluster are recovered together, while the inter-cluster interference is only partially recovered to enhance the intra-cell messages recovery.

This can be thought of as a multiple-input multiple-output (MIMO) MAC setting, where each transmitter has one antenna,
and the receiver is equipped with seven antennas.

Note that in this case there is an asymmetry between the middle cell and its surrounding six neighboring cells within the same cluster.

In the one-dimensional counterpart of this approach, depicted in \figref{fig:HK:1D}, each cluster is composed of a cell triplet. Within each cluster, the messages of the cells comprising the cluster are fully recovered.
The observed channel by the cluster, composed of cells $\{ -1,0,1 \}$, is
\begin{align}
	\by = \bH \bx + \bHinter \bxinter + \bz ,
\end{align}
where the channel output, input, noise and intercell interference vectors are defined as
\begin{align}
\label{eq:katz:vectors}
\!\!\!
\!\!\!
	\by =
	\begin{bmatrix}
		y_{-1}
	 \\ 	y_0
	 \\	y_1
	\end{bmatrix}
	,\,
	\bx =
	\begin{bmatrix}
		x_{-1}
	 \\ 	x_0
	 \\	x_1
	\end{bmatrix}
	,\,
	\bz =
	\begin{bmatrix}
		z_{-1}
	 \\ 	z_0
	 \\	z_1
	\end{bmatrix}
	,\,
	\bxinter =
	\begin{bmatrix}
		x_{-2}
	 \\	x_2
	\end{bmatrix}
	,
\end{align}
and the channel and intercell interference matrices are 
\begin{align}
\label{eq:katz:matrices}
 	\bH &=
 	\begin{bmatrix}
 		1 & \alpha & 0
 	 \\	\alpha & 1 & \alpha
 	 \\ 	0 & \alpha & 1
 	\end{bmatrix}
	,
	&
	\bHinter &=
 	\begin{bmatrix}
 		\alpha & 0
 	 \\ 	0 & 0
 	 \\	0 & \alpha
 	\end{bmatrix}
 	.
\end{align}

The na\"ive approach in this case would be to treat  $\bxinter$ as Gaussian noise. The corresponding (average per user) achievable rate in this case, which is equal in turn to one-third of the sum-capacity of the resulting Gaussian MIMO MAC is equal to\footnote{We use all the power available in this case for all the cells, even though it might be suboptimal in general as cells $\pm 1$ act also as interference for adjacent clusters.}
\begin{align}
	R = \frac{1}{3} \times \frac{1}{2} \log \frac{\left| \bI + P \bHinter \bHinter^T + P \bH \bH^T \right|}{\left| \bI + P \bHinter \bHinter^T \right|}
	\:.
\end{align}

We next describe the one-dimensional version of the scheme of Katz \etal~\cite{KatzZaidelShamai:ISIT2013}.
According to this approach, the signal transmitted by each non-central (``peripheral'') cell $x_{\pm 1}$ is split into two parts,
denoted by $x_i^d$ and $x_i^{ud}$, where `$d$' and `$ud$' stand for `decode' and `undecode', respectively:
\begin{align}
\label{eq:xd+xud}
	x_i &= x_i^d + x_i^{ud} \,, & i = 1, 2.
\end{align}
To facilitate the recovery of the messages of the cluster composed of cells $\{ -1, 0, 1\}$, \viz\ $x_0$,
$x_{\pm 1}^d$ and $x_{\pm 1}^{ud}$, we decode, in addition, also $x_{\pm 2}^d$.
Hence, we arrive at the following equivalent MIMO MAC:
\begin{align}
\label{eq:MAC:d_and_ud}
	\bH_d \bx^d + \bH_{ud} \bx^{ud} + \bz ,
\end{align}
where $\by$ and $\bz$ are as in \eqref{eq:katz:vectors}, the vector of `decoded' signals $\bx^d$ is
\begin{align}
	\bx^d &=
	\begin{bmatrix}
		x_{-2}^d
	 & x_{-1}^d
	 & x_{-1}^{ud}
	 & 	x_0
	 &	x_1^{ud}
	 &	x_1^d
	 &	x_2^d
	\end{bmatrix}^T
	,
\end{align}
and the vector of `undecoded' signals which are treated as noise is defined as
\begin{align}
	\bx^{ud} &=
	\begin{bmatrix}
		x_{-2}^{ud}
	 &	x_2^{ud}
	\end{bmatrix}^T
	.
\end{align}
The corresponding matrices are
\begin{align}
 	&\bH_d =
	\begin{bmatrix}
 		\alpha \sqrt{\lambda} & \sqrt{\lambda} & \sqrt{1 - \lambda} & \alpha & 0 & 0 & 0
 	 \\	0 & \alpha \sqrt{\lambda} & \alpha \sqrt{1 - \lambda} & 1 & \alpha \sqrt{1 - \lambda} & \alpha \sqrt{\lambda} & 0
 	 \\ 	0 & 0 & 0 & \alpha & \sqrt{1 - \lambda} & \sqrt{\lambda} & \alpha \sqrt{\lambda}
 	\end{bmatrix}
 	,
 	\\
 	&\bH_{ud} = \sqrt{1 - \lambda} \bHinter \,.
\end{align}
The (average per user) achievable rate is therefore
\begin{align}
	R = \max_{\lambda \in [0, 1]} \max_{R^d, R^{ud}} \frac{1}{3} \left( R_0 + 2 R_d + 2 R_{ud}  \right)
\end{align}
where
\begin{align}
\begin{aligned}
	&\left( R^d, R^d, R^{ud}, R_0, R^{ud}, R^d, R^d \right) \in
  \col{}{\\ &\qquad\qquad\qquad\qquad\qquad\quad} \Cmac{\bH_d, \bI + P \bH_{ud} \bH_{ud}^T} .
\end{aligned}
\end{align}


\section{Overlapped Clustering}
\label{s:overlap}

The asymmetric nature of the non-overlapping clustering technique of \secref{s:HK} suggests an inherent loss.
To overcome this loss, we propose in this section a symmetrization of this technique by allowing cluster \emph{overlapping}. Moreover, we demonstrate that a sub-optimal variant of the overlapping clustering scheme with \emph{reduced complexity} achieves near optimal performance for practical SNR and $\alpha$ values.

To that end, consider the following scheme: \textit{Each} cell uses the received signals of its neighboring cells, in addition to its own received signal, for the recovery of its message.
However, in contrast to the ``non-overlapping cluster'' approach, the messages of the neighboring cells need not be fully recovered,
and partial recovery can be used instead. This gives an extra degree of freedom, which enhances performance.

\begin{remark}
    Compared to the approach of \secref{s:HK}, 
    the required backhaul communication between BSs is seven times larger, 
    since every BS has to receive the (analog) outputs received by each neighboring BS. 
    Nonetheless, the additional backhaul resources allow to reduce the complexity of the scheme and to improve substantially the achievable rates (simultaneously).
    Note further, that under the alternative C-RAN paradigm, the required backhaul resources are the same.
\end{remark}

In contrast to the non-overlapped scheme of \secref{s:HK} where $x_{\pm 1}$ had to be \emph{fully recovered}, 
in the overlapped variant, there is no requirement to fully recover $x_{\pm 1}$, and partial recovery of these signals can be preferred instead.
A full description of this scheme is available in the appendix.
We now, we concentrate on a simpler variant in which the signals outside the cluster, $x_{\pm 2}$, are treated as noise. Interestingly, even this suboptimal variant gives substantial improvement for a wide range of parameters over the non-overlapped clustering scheme of \secref{s:HK}.

The simplified variant for the one-dimensional case works as follows.
Each signal $x_i$ is divided into two parts as in \eqref{eq:xd+xud}, and the resulting MAC is described by
\begin{align}
\label{eq:overlap_cluster:MAC}
    \by = \bH_d \bx^d + \bH_{ud} \bx^{ud} + \bHinter \bxinter + \bz ,
\end{align}
where $\by$, $\bz$ and $\bxinter$ are defined as in \eqref{eq:katz:vectors},
and
\begin{align}
	\bx^d &=
	\begin{bmatrix}
	    x_{-1}^d
	 & 	x_0^d
	 & 	x_0^{ud}
	 &	x_1^d
	\end{bmatrix}^T
	\\
	\bx^{ud} &=
	\begin{bmatrix}
	     x_{-1}^{ud}
	 &	x_1^{ud}
	\end{bmatrix}^T
	.
\end{align}
$\bHinter$ is given in \eqref{eq:katz:matrices},
and $\bH_d$ and $\bH_{ud}$ are defined as
\begin{subequations}
\label{eq:tetris:Hs}
\noeqref{eq:tetris:Hs:Hd,eq:tetris:Hs:Hud}
\begin{align}
	\bH_d &=
	\begin{bmatrix}
		\sqrt{\lambda} & \alpha \sqrt{\lambda} & \alpha \sqrt{1 - \lambda} & 0
	 \\  \alpha \sqrt{\lambda} & \sqrt{\lambda} & \sqrt{1 - \lambda} & \alpha \sqrt{\lambda}
	 \\  0 & \alpha \sqrt{\lambda} & \alpha \sqrt{1 - \lambda} & \sqrt{\lambda}
	\end{bmatrix}
\label{eq:tetris:Hs:Hd}
\\
   \bH_{ud} &= \sqrt{1 - \lambda}
 	\begin{bmatrix}
 		1 & 0
 	 \\  \alpha & \alpha
 	 \\  0 & 1
 	\end{bmatrix}
 	.
\label{eq:tetris:Hs:Hud}
\end{align}
\end{subequations}
The achievable rate, using this technique is given by
\begin{align}
	R = \max_{\lambda \in [0,1]} \max_{R^d, R^{ud}} \left\{ R^d + R^{ud} \right\} ,
\end{align}
where 
\begin{align}
\begin{aligned}
	\left( R^d, R^d, R^{ud}, R^d  \right) &\in
	\col{}{\\ &\!\!\!\!\!\!\!\!\!\!\!\!\!\!\!\!\!\!\!}
		\Cmac{\bH_d, \bI + P \bH_{ud} \bH_{ud}^T + P \bHinter \bHinter^T} .
\end{aligned}
\end{align}


\section{Time Sharing}
\label{s:time-sharing}

The simplest scheme that involves scheduling is that of time sharing. 
A two-phase time-sharing scheme alternates between transmission of odd-indexed cells during odd time instants, 
and transmission of even-indexed cells during even time instants. Since every cell transmits only during half of the time, 
it can utilize twice the total available power during its transmission periods.
The achievable rate when using this scheme is 
\begin{align}
    R_\text{TS} = \frac{1}{2} \C{2 P} .
\end{align}


\section{Scheduled Decoding: Digital Backhaul}
\label{s:tetris:1antenna}

In this section we consider the use of only digital backhaul communication between adjacent cells.
Such backhaul communication is put to use by adopting a multi-stage decoding scheduling policy.
This allows to remove interference from neighboring signals
that were already recovered by their respective cells, stage-by-stage.

For the one-dimensional case, we propose a two-phase schedule:
First only cells with even indices recover their respective messages, treating the interfering messages of the adjacent (odd-indexed) cells as noise.
Thus, the rate recovered by these cells is equal to \eqref{eq:naive_rate}.
The even-indexed cells convey their recovered messages to their neighboring odd-indexed cells. The odd-indexed cells can now subtract the interfering even-indexed signals prior to the recovery of the (odd-indexed) signals intended to them:
\begin{subequations}
\begin{align}
    y^\text{Sch}_{2i+1} &= y_{2i+1} - \alpha \left( x_{2i} - x_{2i+2} \right)
 \\ &= x_{2i+1} + z_{2i+1} .
\end{align}
\end{subequations}
Hence, the rate of these cells is equal to $\C{P}$.

The average achievable rate is therefore equal to
\begin{align}
	R_\text{Sch} = \frac{1}{2} \left[ \C{ \frac{P}{1 + 2 \alpha^2 P} } + \C{P} \right] .
\end{align}

This scheme clearly outperforms its non-scheduled counterpart \eqref{eq:naive_rate}. Interestingly, it outperforms the non-overlapping and (simplified) overlapping clustering schemes of \secref{s:HK} and \secref{s:overlap}, for some channel parameters.
Thus, incorporating scheduled decoding into the clustered scheme of \secref{s:overlap} promises substantial improvement. This is discussed in the next section.


\section{Cluster Overlap with Scheduled Decoding}
\label{s:tetris}

In this section we combine the (simplified) overlapping clustering technique of \secref{s:overlap} with that of scheduled decoding of \secref{s:tetris:1antenna}. Namely, we make use of both analog and digital backhaul communication.

We note that, in contrast to \secref{s:tetris:1antenna} where the desired message was recovered from the received signal of a single cell, since in \secref{s:overlap} the received signals of three adjacent cells (``cluster'') are used,
adding a third and a forth phase to the schedule promises further enhancement in performance.
Nonetheless, we shall describe the two-phase scheme; the extension to a four-phase scheme is straightforward.

In the two-phase scheme, as in \secref{s:tetris:1antenna}, first the even-indexed messages are recovered by their respective cells, each from its received vector $\by$, which is defined as in \eqref{eq:katz:vectors}.
Moreover, as in \secref{s:overlap} (and \secref{s:HK}),
the odd-indexed transmitted signals are composed of two parts, $\{x_{2i+1}^d\}$ and $\{x_{2i+1}^{ud}\}$,
which correspond to the parts that are recovered during the first (``even'') and second (``odd'') phases, respectively.\footnote{Since the even-indexed signals are recovered \emph{entirely} during the first phase, there is no need to split them into two parts as well.}
Hence, during the first phase, in addition to recovering $\left\{ x_{2i} \right\}$,
also $\left\{ x_{2i+1}^d \right\}$ are recovered.
The corresponding MIMO MAC describing the first phase is 
\begin{align}
	\by = \bH_d \bx^d + \bH_{ud} \bx^{ud} + \bHinter \bxinter + \bz ,
\end{align}
where the vectors are defined as
\begin{align}
	\by &\triangleq
	\begin{bmatrix}
	     y_{2i-1}
	 \\  y_{2i}
	 \\  y_{2i+1}
	\end{bmatrix}
	,&
	\bx^{d} &\triangleq
	\begin{bmatrix}
	     x_{2i-1}^d
	 \\  x_{2i}
	 \\  x_{2i+1}^d
	\end{bmatrix}
	,&
	\bz &\triangleq
	\begin{bmatrix}
	     z_{2i-1}
	 \\  z_{2i}
	 \\  z_{2i+1}
	\end{bmatrix}
	\col{,&}{\\}
	\bx^{ud} &\triangleq
	\begin{bmatrix}
	     x_{2i-1}^{ud}
	 \\  x_{2i+1}^{ud}
	\end{bmatrix}
	,&
	\bxinter &\triangleq
	\begin{bmatrix}
	     x_{2i-2}^{ud}
	 \\  x_{2i+2}^{ud}
	\end{bmatrix}
	,
\end{align}
\noindent
the matrix $\bH_{ud}$ is given in \eqref{eq:tetris:Hs:Hud},
and $\bHinter$ and $\bH_d$ are defined here as
\begin{align}
	\bH_d &=
	\begin{bmatrix}
		\sqrt{\lambda} & \alpha & 0
 	 \\  \alpha \sqrt{\lambda} & 1 & \alpha \sqrt{\lambda}
 	 \\  0 & \alpha & \sqrt{\lambda}
	\end{bmatrix}
	, &
	\bHinter &= \sqrt{1 - \lambda}
 	\begin{bmatrix}
 		\alpha & 0
 	 \\ 	0 & 0
 	 \\	0 & \alpha
 	\end{bmatrix}
 	.
\end{align}

The corresponding rates $R_\text{odd}^d$ and $R_\text{even}$ need to satisfy
\begin{align}
	& \left(R_\text{odd}^d, R_\text{even}, R_\text{odd}^d \right) \in
 \col{}{\\  & \qquad\qquad\quad} \Cmac{\bH_d, \bI + P \bHinter \bHinter^T + P \bH_{ud} \bH_{ud}^T} ,
\end{align}
where $R_\text{even}$ is the rate of $\{x_{2i}\}$ and $R_\text{odd}^d$ is the rate of $\{x_{2i-1}^d\}$.

In the second phase, the interference of the recovered signals during the first phase is subtracted, after which the remaining messages $\left\{ x_{2i+1}^{ud} \right\}$ are recovered. The corresponding MAC in this case is
\begin{align*}
	\begin{bmatrix}
	     y_{2i}
	 \\  y_{2i+1}
	 \\  y_{2i+2}
	\end{bmatrix}
	&=
	\sqrt{1 - \lambda}
	\left(
	\begin{bmatrix}
		\alpha
	 \\  1
	 \\  \alpha
	\end{bmatrix}
	x^{ud}_{2i+1}
	+
	\alpha
	\begin{bmatrix}
		x^{ud}_{2i-1}
 	 \\  0
 	 \\  x^{ud}_{2i+1}
	\end{bmatrix}
	\right)
	+
	\begin{bmatrix}
		z_{2i}
	 \\  z_{2i+1}
	 \\  z_{2i+2}
	\end{bmatrix}
\end{align*}
where the intercluster signals $x^{ud}_{2i \pm 1}$ are treated as noise.
Thus, the achievable rate during the second phase is equal to
\begin{align}
	R_\text{odd}^{ud} = \C{ (1-\lambda) P \left[ 1 + \frac{2 \alpha^2}{1 + 1(1-\lambda) \alpha^2 P} \right] } ,
\end{align}
and hence the average rate per cell of the scheme is equal to
\begin{align}
	R = \frac{1}{2} \max_{\lambda \in [0, 1]} \left\{ R_\text{even} + R_\text{odd}^d + R_\text{odd}^{ud} \right\} .
\end{align}


\begin{figure}[t]
	\includegraphics[width=\columnwidth]{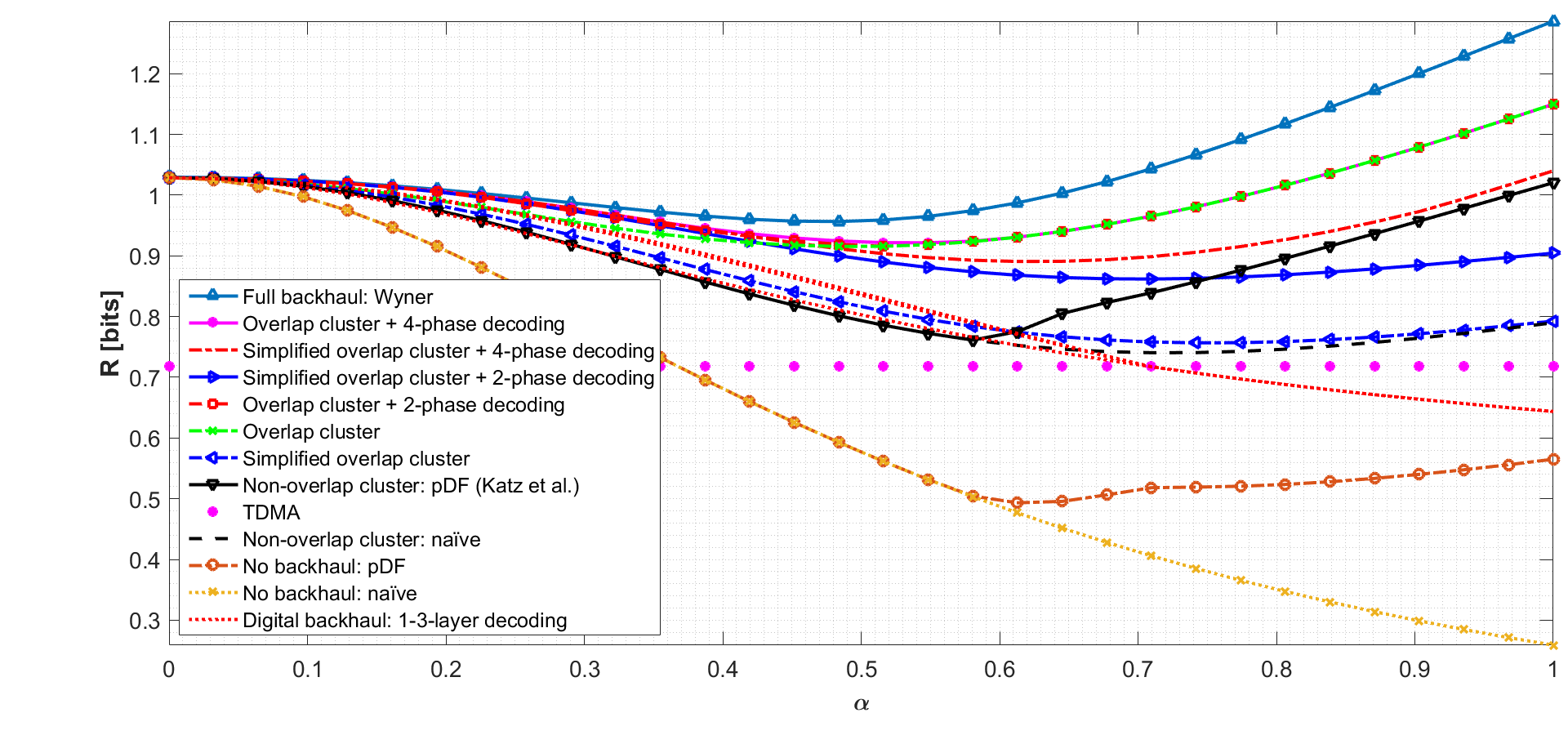}
	\caption{Achievable rates over the one-dimensional array for $\SNR =5\, \dB$.}
\label{fig:rates:1D:5dB}
\end{figure}
\begin{figure}[t]
	\includegraphics[width=\columnwidth]{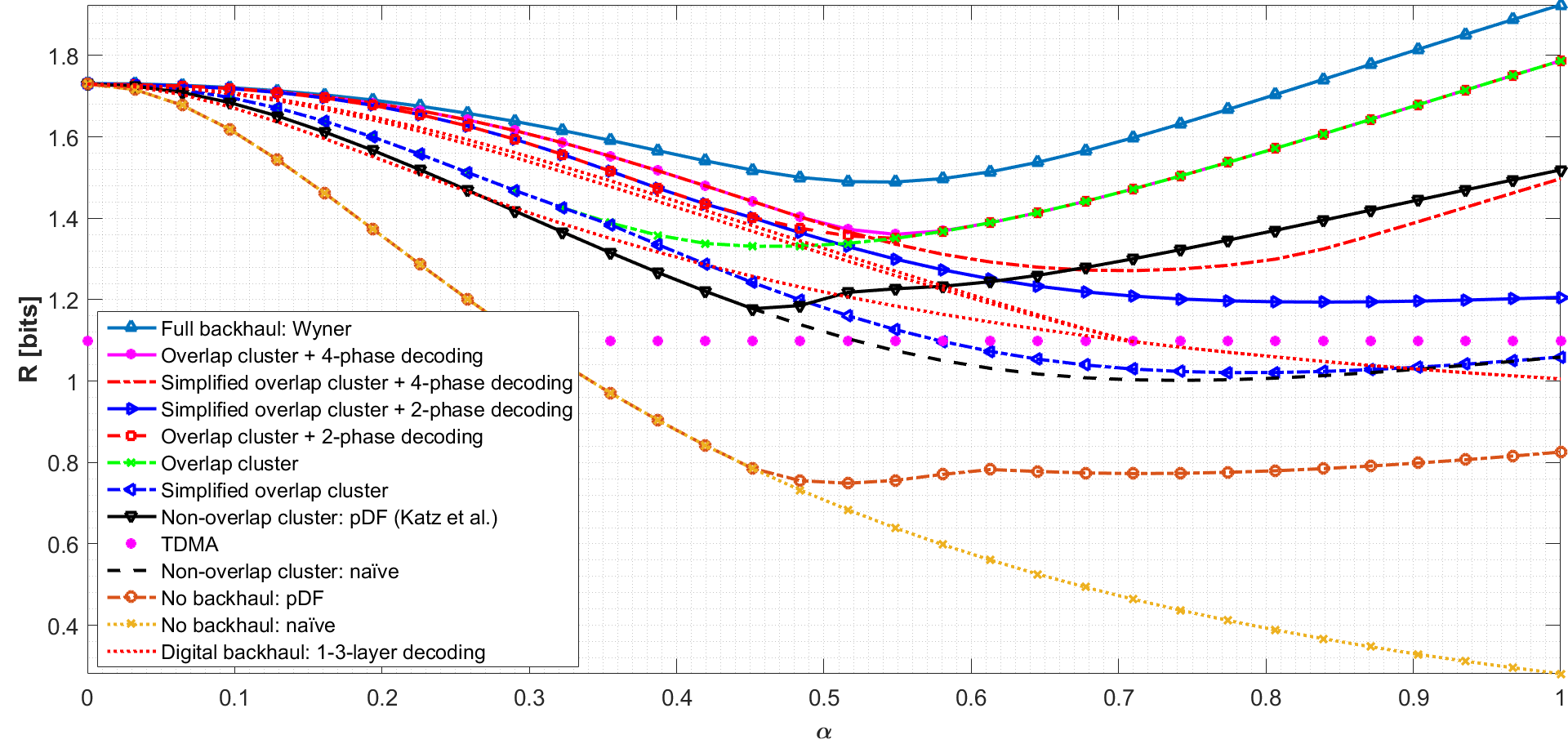}
	\caption{Achievable rates over the one-dimensional array for $\SNR =10\, \dB$.}
\label{fig:rates:1D:10dB}
\end{figure}
\begin{figure}[t]
	\includegraphics[width=1\columnwidth]{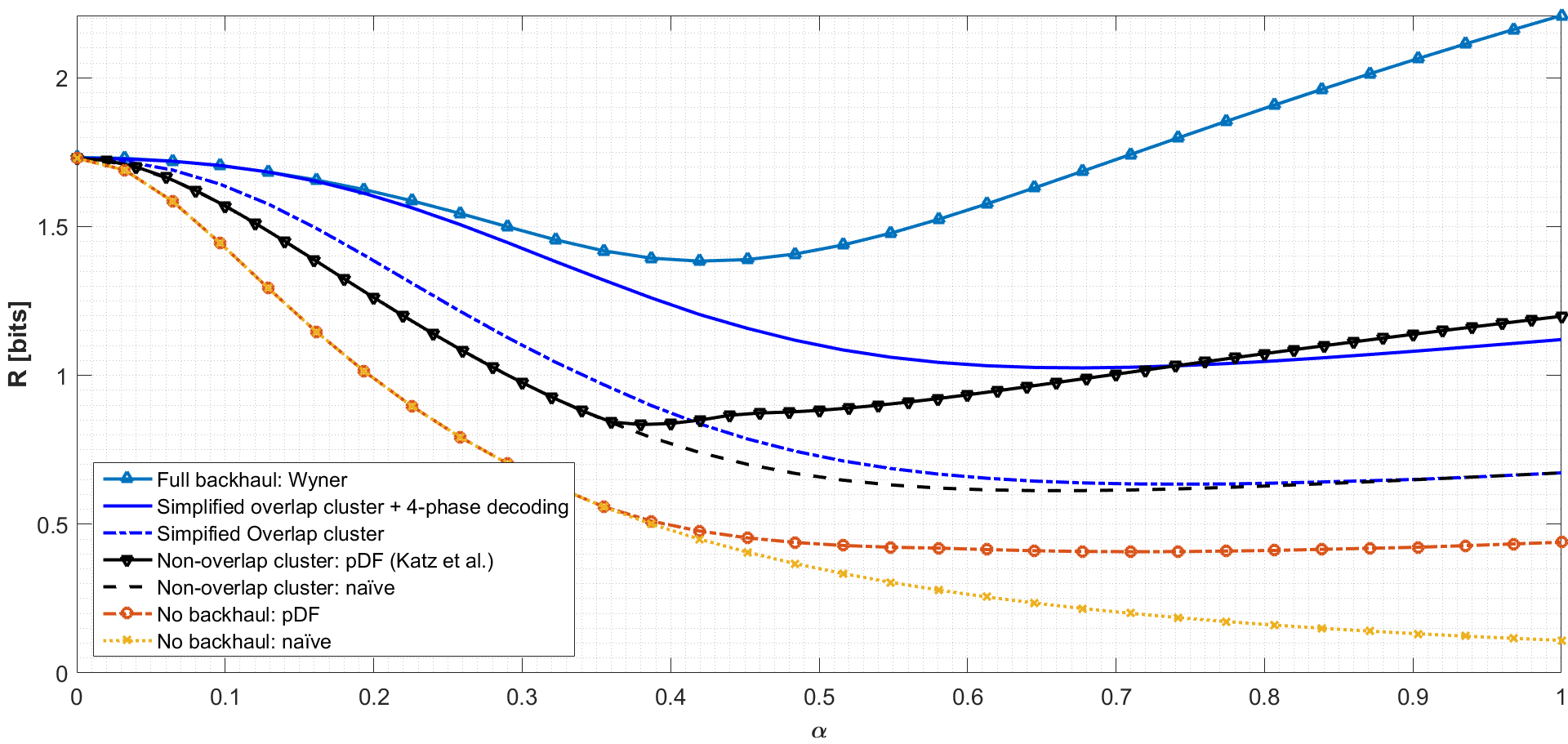}
	\caption{Achievable rates over the two-dimensional array for $\SNR =10\, \dB$.}
\label{fig:rates:2D:10dB}
\end{figure}

\section{Multi-layered Scheduled Decoding}
\label{s:moses}

In this section we improve upon the scheduled decoding scheme of \secref{s:tetris}, by breaking each transmitted signal into multiple layers and recovering them over multiple decoding phases.

To understand the added value of such a multi-layered structure, we note that each signal serves both as the desired/target signal for its respective cell and as interference for its neighboring cells.
By breaking each message into multiple layers, and recover the messages of even- and odd-indexed cells
alternately, the interference experienced by the even-indexed cells is effectively reduced
and the rates achievable by the two groups of cells are better balanced.

Alternatively, this scheme can be thought of as a poor man's version of Wyner's joint processing scheme of \secref{s:Wyner},
in which a message-passing like algorithm is materialized via the incremental layer recovery and their exchange between adjacent cells.

\begin{remark}
    Over the point-to-point AWGN channel, a layered scheme that breaks each message into sub-messages
    and encodes each sub-message into a different code layer, 
    attains capacity for any
    power allocation between the layers and codes of appropriate rates. Moreover it can be achieved under successive interference cancelation,
    where at each step previously recovered layers are subtracted and non-decoded yet layers are treated as noise for the recovery of the current layer \cite{RimoldiUrbanke:RateSplitting}.
\end{remark}

We shall demonstrate the idea behind this scheme for the case where only ``digital backhaul'' is allowed, \ie,
when each cell can use its own received signal along with the messages recovered by its neighboring cells.
\second{Interestingly, this scheme outperforms all previously discussed schemes for certain $\alpha$ and SNR parameters, as is demonstrated in \figref{fig:rates:1D:25dB}.}{}

Denote by $M$ the number of code layers used in each cell,
and by $x_i^{(\ell)}$~--- the $\ell$-th code layer ($\ell = 1,  \ldots, M$) in cell $i$. We shall assume that each such code layer is of power $P$.
The MAC observed by an even-indexed cell, say $i = 0$ is
\begin{align}
\label{eq:Moses:MAC:even}
\begin{aligned}
	y_0 = \sum_{\ell = 1}^M \sqrt{\lambda_\text{even}^{(\ell)}} x_0^{(\ell)}
	&+ \alpha \sum_{\ell = 1}^M \sqrt{\lambda_\text{odd}^{(\ell)}} x_{-1}^{(\ell)}
 \col{}{\\*  &}+ \alpha \sum_{\ell = 1}^M \sqrt{\lambda_\text{odd}^{(\ell)}}x_{1}^{(\ell)}
	+ z_0 \,,
\end{aligned}
\end{align}
and the MAC observed by an odd-indexed cell, say $i = 1$, is
\begin{align}
\label{eq:Moses:MAC:odd}
\begin{aligned}
	y_1 = \sum_{\ell = 1}^M \sqrt{\lambda_\text{odd}^{(\ell)}} x_1^{(\ell)}
	&+ \alpha \sum_{\ell = 1}^M \sqrt{\lambda_\text{even}^{(\ell)}} x_{0}^{(\ell)}
 \col{}{\\*  &}+ \alpha \sum_{\ell = 1}^M \sqrt{\lambda_\text{even}^{(\ell)}}x_{2}^{(\ell)}
	+ z_1 \,,
\end{aligned}
\end{align}
where $\lambda_\text{even}^{(\ell)}$ and $\lambda_\text{odd}^{(\ell)}$ are
the power portions allocated to layer $\ell$ in even-indexed cells and in odd-indexed cells, respectively,
and therefore satisfy
\begin{align}
	\sum_{\ell = 1}^M \lambda_\text{even}^{(\ell)} = \sum_{\ell = 1}^M \lambda_\text{odd}^{(\ell)} = 1 .
\end{align}

We assume correct decoding of previously decoded signals, \ie, that $\{x_i^{(\ell)} |\, \ell = 1, \ldots, k-1;\, i = -1,0,1 \}$  are known when recovering $x_0^{(k)}$,
and $\{x_i^{(\ell)} |\, \ell = 1, \ldots, k;\, i = 0,2 \}$ and $\{x_1^{(\ell)} |\, \ell = 1, \ldots, k-1 \}$ are known when recovering $x_1^{(k)}$.

Thus, after subtracting the components of the previously recovered messages, $x_0^{(k)}$ is recovered from the following channel:
\begin{subequations}
\label{eq:Moses:MAC:even}
\noeqref{eq:Moses:MAC:even:signal,eq:Moses:MAC:even:ISI}
\begin{align}
    y_0^{(k)} &=
    y_0 - \sum_{\ell = 1}^{k-1} \sqrt{\lambda_\text{even}^{(\ell)}} x_0^{(\ell)}
    - \alpha \sum_{\ell = 1}^{k-1} \sqrt{\lambda_\text{odd}^{(\ell)}} x_{-1}^{(\ell)}
\col{}{\nonumber
 \\ &\qquad \qquad \qquad \qquad \quad \:\:\,} - \alpha \sum_{\ell = 1}^{k-1} \sqrt{\lambda_\text{odd}^{(\ell)}} x_1^{(\ell)}
 \\ &= \sqrt{\lambda_\text{even}^{(k)}} x_0^{(k)} + z_0
+ \sum_{\ell = k+1}^M \sqrt{\lambda_\text{even}^{(\ell)}} x_0^{(\ell)}
\col{}{\nonumber
\\* &\quad} + \alpha \sum_{\ell = k}^M \sqrt{\lambda_\text{odd}^{(\ell)}} x_{-1}^{(\ell)}
 + \alpha \sum_{\ell = k}^M \sqrt{\lambda_\text{odd}^{(\ell)}}x_{1}^{(\ell)}
\label{eq:Moses:MAC:even:ISI:inter}
	\,,
\end{align}
\end{subequations}
where the other code layers in \eqref{eq:Moses:MAC:even} are treated as noise.

The achievable rate in recovering $x_0^{(k)}$ is therefore
\begin{align}
	R_0^{(k)} = \C{ \frac{\lambda_\text{even}^{(k)} P}{1 + \sum_{\ell = k+1}^M \lambda_\text{even}^{(\ell)} P + 2 \sum_{\ell = k}^M \lambda_\text{odd}^{(\ell)} P } } .
\end{align}

Similarly, $x_1^{(k)}$ is recovered from the channel
\begin{subequations}
\label{eq:Moses:MAC:odd}
\noeqref{eq:Moses:MAC:odd:signal,eq:Moses:MAC:odd:ISI}
\begin{align}
    y_1^{(k)} &=
    y_1 - \sum_{\ell = 1}^{k-1} \sqrt{\lambda_\text{odd}^{(\ell)}} x_1^{(\ell)}
    - \alpha \sum_{\ell = 1}^k \sqrt{\lambda_\text{even}^{(\ell)}} x_{0}^{(\ell)}
\col{}{\nonumber
 \\ &\qquad \qquad \qquad \qquad \quad \:\;} - \alpha \sum_{\ell = 1}^k \sqrt{\lambda_\text{even}^{(\ell)}} x_{2}^{(\ell)}
 \\ &= \sqrt{\lambda_\text{odd}^{(k)}} x_1^{(k)} + z_1
    + \sum_{\ell = k+1}^M \sqrt{\lambda_\text{odd}^{(\ell)}} x_1^{(\ell)}
\col{}{\nonumber
    \\* &\quad} + \alpha \sum_{\ell = k+1}^M \sqrt{\lambda_\text{even}^{(\ell)}} x_{0}^{(\ell)}
     + \alpha \sum_{\ell = k+1}^M \sqrt{\lambda_\text{even}^{(\ell)}}x_{2}^{(\ell)}
\label{eq:Moses:MAC:odd:ISI:inter}
	\,,
\end{align}
\end{subequations}
and the corresponding achievable rate in recovering $x_1^{(k)}$ is
\begin{align}
	R_1^{(k)} = \C{ \frac{\lambda_\text{even}^{(k)} P}{1 + \sum_{\ell = k+1}^M \left( \lambda_\text{even}^{(\ell)} P + 2 \lambda_\text{odd}^{(\ell)} P \right) } } .
\end{align}

\begin{figure*}[t]
\begin{subequations}
\label{eq:tetris_full:Hs}
\noeqref{eq:tetris_full:Hs:Hd,eq:tetris_full:Hs:Hud}
\begin{align}
    \bH_d &=
    \begin{bmatrix}
	\alpha \sqrt{\lambda_\inter} & \sqrt{\lambda_\inter}	& \sqrt{\lambda_\intra}		& \alpha \sqrt{\lambda_\inter} & \alpha \sqrt{\lambda_\intra} & \alpha \sqrt{\lambda_\self} & 0 & 0 & 0
     \\ 0				 & \alpha \sqrt{\lambda_\inter}	& \alpha \sqrt{\lambda_\intra}	& \sqrt{\lambda_\inter}	&	\sqrt{\lambda_\intra} & \sqrt{\lambda_\self} & \alpha \sqrt{\lambda_\inter} & \alpha \sqrt{\lambda_\intra} & 0
     \\ 0				 & 0				& 0				& \alpha \sqrt{\lambda_\inter} & \alpha \sqrt{\lambda_\intra} & \alpha \sqrt{\lambda_\self} & \sqrt{\lambda_\inter} & \sqrt{\lambda_\intra} & \alpha \sqrt{\lambda_\inter}
    \end{bmatrix}
\label{eq:tetris_full:Hs:Hd}
\\[.5\baselineskip]
   \bH_{ud} &= 
    \begin{bmatrix}
	\alpha \sqrt{\lambda_\intra}	& \alpha \sqrt{\lambda_\self} 	& \sqrt{\lambda_\self}		& 0				& 0				     & 0
     \\ 0				& 0				& \alpha \sqrt{\lambda_\self}	& \alpha \sqrt{\lambda_\self}	& 0					& 0
     \\ 0				& 0				& 0				& \sqrt{\lambda_\self}		& \alpha \sqrt{\lambda_\intra}	& \alpha \sqrt{\lambda_\self}
    \end{bmatrix}
    .
\label{eq:tetris_full:Hs:Hud}
\end{align}
\end{subequations}
\hrulefill
\end{figure*}
The achievable rate is therefore 
\begin{align}
    R = \max_{\lambda_\self, \lambda_\inter, \lambda_\intra} 
	\,
        \max_{R^\self, R^\inter, R^\intra} R^\self + R^\inter + R^\intra
        ,
\end{align}
where
\begin{align}
    &\left( R^\inter, R^\inter, R^\intra, R^\inter, R^\intra, R^\self, R^\inter, R^\intra, R^\inter \right) 
 \col{}{\\ &\qquad\qquad\qquad\qquad\qquad\qquad} \in \Cmac{ \bH_d ,\, \bI + P \bH_{ud} \bH_{ud}^T}
    .
\end{align}


\section{Performance Comparison}
\label{s:numeric}

\begin{figure}[t]
	\includegraphics[width=1\columnwidth]{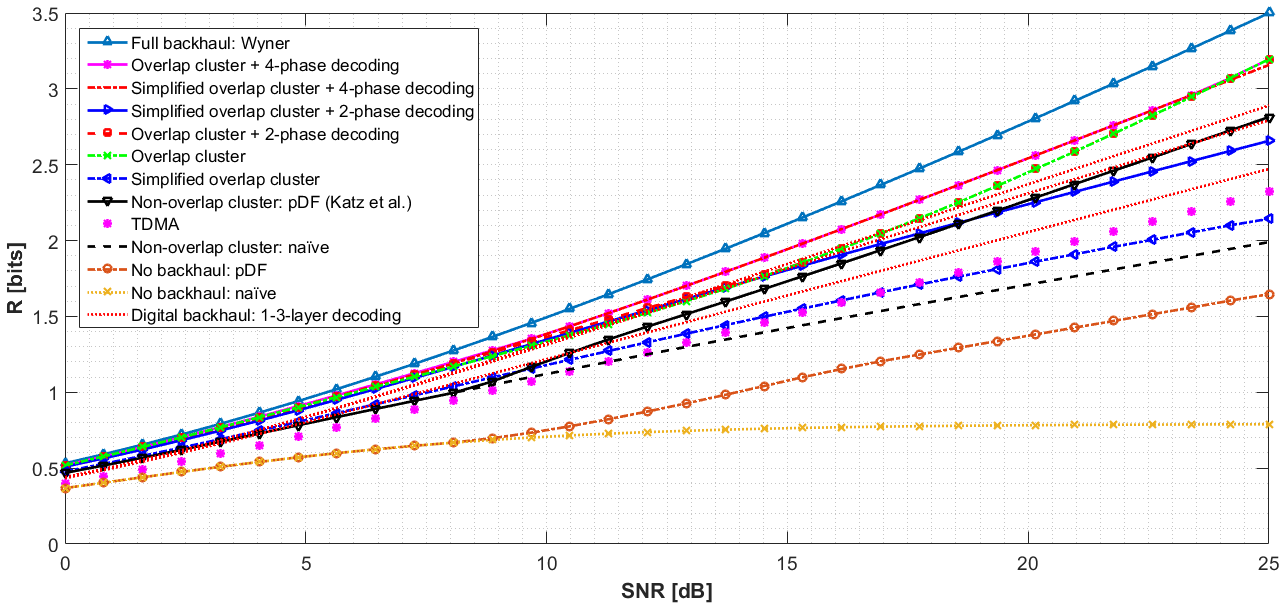}
	\caption{Achievable rates over the one-dimensional array for $\alpha = 0.5$.}
\label{fig:rates:1D:alpha=0.5}
\end{figure}

\begin{figure}[t]
	\includegraphics[width=1\columnwidth]{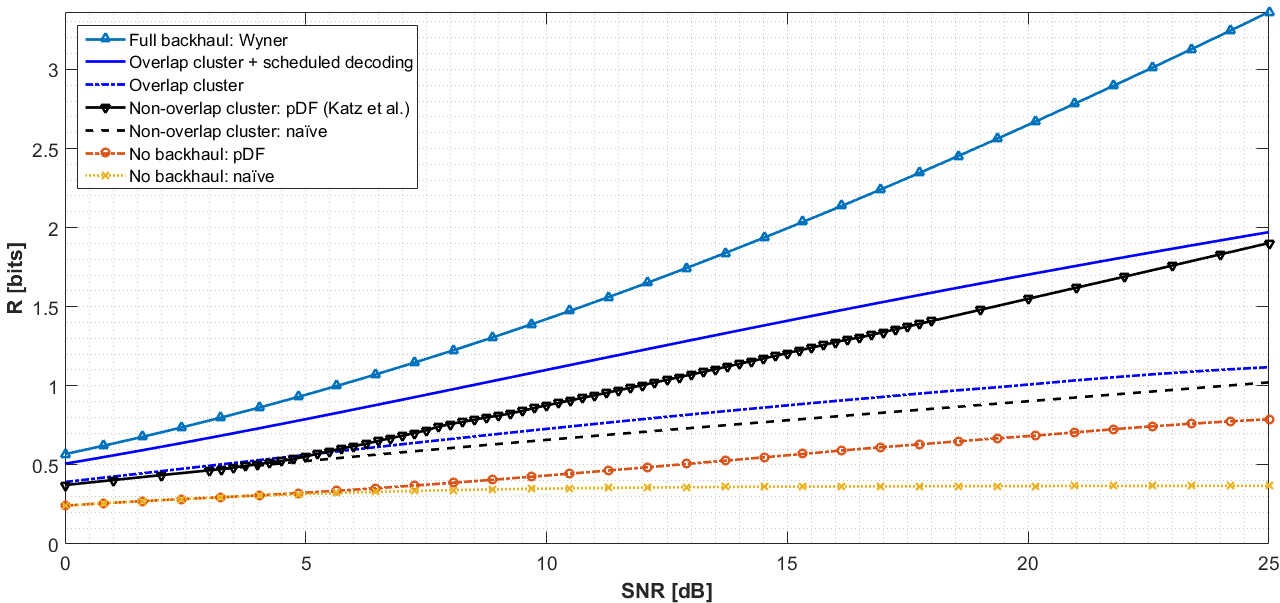}
	\caption{Achievable rates over the two-dimensional array for $\alpha = 0.5$.}
\label{fig:rates:2D:alpha=0.5}
\end{figure}

    In this section we compare the performance of the different schemes discussed in Sections \ref{s:Wyner}--\ref{s:moses}
    for both the one- and two-dimensional models, at $\SNR = 10 \dB$ and $\alpha \in [0,1]$; the results are depicted in Figures \ref{fig:rates:1D:10dB} and \ref{fig:rates:2D:10dB}, and Figures \ref{fig:rates:1D:alpha=0.5} and \ref{fig:rates:2D:alpha=0.5}, 
    respectively.

    It is apparent from Figures \ref{fig:rates:1D:5dB}--\ref{fig:rates:1D:alpha=0.5} that 
    the scheme of \secref{s:tetris} achieves close-to-capacity performance for all practical values of $\alpha$ and SNR.
    Furthermore, for low-to-intermediate $\alpha$ values simplified overlapped clustering with 4-phase decoding coincides with 
    the non-simplified variant, suggesting that essentially no partial decoding of the out-of-cluster interference is performed. For intermediate-to-high $\alpha$ values, overlapped clustering without scheduled decoding achieves the same results as if we allowed clustering; this suggests that in this regime scheduling does not enhance performance. Finally, we note that in the two-dimensional regime, the situation is similar: simplified overlapped clustering promises substantial gain over the scheme by Katz \etal, especially when combined with scheduled decoding.
    
	

\section{Discussion: Digital backhaul vs.\ Analog backhaul}
\label{s:discuss}

The scheduled decoding schemes of Sections \ref{s:tetris:1antenna} and \ref{s:tetris}, can be improved by breaking each transmitted signal into multiple layers and recovering them over multiple decoding phases as in \secref{s:moses}.
In fact, for low-to-moderate $\alpha$ values, we observe that digital backhaul in conjunction with multi-layered scheduled decoding 
gives close-to-capacity results. This offers, in turn, a great reduction in the backhaul communication required to attain similar results with analog backhaul.


\section{Acknowledgments}

We thank Gil Katz for providing us the source of Figures~3 and 4 in \cite{KatzZaidelShamai:ISIT2013} and his thesis \cite{katz_msc}.

\appendix[Overlapped Clustered \\ Partial Decode-and-Forward]

In the (non-simplified) partial decode-and-forward scheme for the overlapped cluster, 
each signal $x_i$ is divided into three parts:
\begin{align}
    x_i = x^\self_i + x^\intra_i + x^\inter_i .
\end{align}
The resulting MAC is given as 
\begin{align}
    \by = \bH_d \bx^d + \bH_{ud} \bx^{ud} + \bz , 
\end{align} 
with $\by$, $\bz$ defined as in \eqref{eq:katz:vectors},
and 
\begin{align}
\col{}{	&
\!\!}
	\left( \bx^d \right)^T \col{&}{}=
\col{}{    \\  &
\!\!}
	\begin{bmatrix}
		x_{-2}^\inter
	 &  	x_{-1}^\inter
	 & 	x_{-1}^\intra
	 & 	x_0^\inter
	 & 	x_0^\intra
	 & 	x_0^\self
	 &	x_1^\inter
	 &	x_1^\intra
	 &	x_2^\inter
	\end{bmatrix}
	\\
\col{}{	&
\!\!}
	\left( \bx^{ud} \right)^T \col{&}{}=
	\begin{bmatrix}
		x_{-2}^\intra
	 &	x_{-2}^\self
	 &	x_{-1}^\self
	 &	x_{1}^\self
	 &	x_{2}^\intra
	 &	x_{2}^\self
	\end{bmatrix}
	;
\end{align}
the effective channel matrices are given in~\eqref{eq:tetris_full:Hs} where 
$\lambda_\self$, $\lambda_\inter$ and $\lambda_\intra$ are the power portions allocated to $x^\self_i$, $x^\inter_i$ and $x^\intra_i$, respectively, and therefore are all non-negative and satisfy $\lambda_\self + \lambda_\inter + \lambda_\intra = 1$.




\begin{thebibliography}{10}
\providecommand{\url}[1]{#1}
\csname url@samestyle\endcsname
\providecommand{\newblock}{\relax}
\providecommand{\bibinfo}[2]{#2}
\providecommand{\BIBentrySTDinterwordspacing}{\spaceskip=0pt\relax}
\providecommand{\BIBentryALTinterwordstretchfactor}{4}
\providecommand{\BIBentryALTinterwordspacing}{\spaceskip=\fontdimen2\font plus
\BIBentryALTinterwordstretchfactor\fontdimen3\font minus
  \fontdimen4\font\relax}
\providecommand{\BIBforeignlanguage}[2]{{%
\expandafter\ifx\csname l@#1\endcsname\relax
\typeout{** WARNING: IEEEtran.bst: No hyphenation pattern has been}%
\typeout{** loaded for the language `#1'. Using the pattern for}%
\typeout{** the default language instead.}%
\else
\language=\csname l@#1\endcsname
\fi
#2}}
\providecommand{\BIBdecl}{\relax}
\BIBdecl

\bibitem{HetNet}
A.~Damnjanovic, J.~Montojo, Y.~Wei, T.~Ji, T.~Luo, M.~Vajapeyam, T.~Yoo,
  O.~Song, and D.~Malladi, ``A survey on {3GPP} heterogeneous networks,''
  \emph{IEEE Trans. Wireless Comm.}, vol.~18, pp. 10--21, 2011.

\bibitem{Survey_ICIC}
A.~S. Hamza, S.~S. Khalifa, H.~S. Hamza, and K.~Elsayed, ``A survey on
  inter-cell interference coordination techniques in {OFDMA}-based cellular
  networks,'' \emph{IEEE Comm.\ Surveys Tutor.}, vol.~15, pp. 1642--1670, 2013.

\bibitem{COMP_3GPP}
D.~Lee, H.~Seo, B.~Clerckx, E.~Hardouin, D.~Mazzarese, S.~Nagata, and
  K.~Sayana, ``Coordinated multipoint transmission and reception in
  lte-advanced: deployment scenarios and operational challenges,'' \emph{IEEE
  Comm.\ Magazine}, vol.~50, no.~2, pp. 148--155, 2012.

\bibitem{Cloud_China_Mobile}
{China Mobile Research Institute}, ``{C-RAN}: The road towards green {RAN},''
  \emph{White Paper, ver. 2.5}, 2011.

\bibitem{CloudRAN_2014_let}
S.~H. Park, O.~Simeone, O.~Sahin, and S.~Shamai, ``Inter-cluster design of
  precoding and fronthaul compression for cloud radio access networks,''
  \emph{IEEE Wireless Comm.\ Let.}, vol.~3, pp. 369--372, 2014.

\bibitem{Poor14_CRAN}
M.~Peng, C.~Wang, V.~K.~N. Lau, and H.~V. Poor, ``Fronthaul-constrained cloud
  radio access networks: Insights and challenges,'' \emph{CoRR}, vol.
  abs/1503.01187, 2015.

\bibitem{Layered_CloudRAN}
S.~H. Park, O.~Simeone, O.~Sahin, and S.~Shamai, ``Robust layered transmission
  and compression for distributed uplink reception in cloud radio access
  networks,'' \emph{IEEE Tran.\ Vehic.\ Tech.}, vol.~63, pp. 204--216, 2014.

\bibitem{CloudRAN_2014}
------, ``Fronthaul compression for cloud radio access networks: Signal
  processing advances inspired by network information theory,'' \emph{IEEE
  Signal Processing Magazine}, vol.~31, pp. 69--79, 2014.

\bibitem{KatzZaidelShamai:ISIT2013}
G.~Katz, B.~M. Zaidel, and S.~Shamai, ``On layered transmission in clustered
  cooperative cellular architectures,'' in \emph{Proc. IEEE Int. Symp.\ on
  Info.\ Theory (ISIT)}, Istanbul, Turkey, July 2013, pp. 1162--1166.

\bibitem{Wyner94}
A.~D. Wyner, ``Shannon-theoretic approach to a {G}aussian cellular
  multiple-access channel,'' \emph{IEEE Trans.\ Inf.\ Theory}, vol.~40, no.~6,
  pp. 1713--1727, Nov. 1994.

\bibitem{HanKobayashi}
T.~S. Han and K.~Kobayashi, ``A new achievable rate region for the interference
  channel,'' \emph{IEEE Trans.\ Inf.\ Theory}, vol.~27, pp. 49--60, 1981.

\bibitem{YuPhD}
W.~Yu, ``Competition and cooperation in multi-user communication
  environments,'' Ph.D. dissertation, Stanford University, 2002.

\bibitem{PiIsWrong_Original}
\BIBentryALTinterwordspacing
R.~Palais, ``$\pi$ is wrong,'' \emph{The Mathematical Intelligencer}, vol.~23,
  no.~3, pp. 7--8, 2001. [Online]. Available: \url{http://bit.ly/pi-is-wrong}
\BIBentrySTDinterwordspacing

\bibitem{PiIsWrong_NoReally}
\BIBentryALTinterwordspacing
M.~Hartl, ``The tau manifesto,'' 2010. [Online]. Available:
  \url{http://tauday.com/tau-manifesto}
\BIBentrySTDinterwordspacing

\bibitem{CoverBook2Edition}
T.~M. Cover and J.~A. Thomas, \emph{Elements of Information Theory, Second
  Edition}.\hskip 1em plus 0.5em minus 0.4em\relax New York: Wiley, 2006.

\bibitem{LeeBook3rd}
J.~Barry, E.~Lee, and D.~Messerschmitt, \emph{Digital Communication}.\hskip 1em
  plus 0.5em minus 0.4em\relax Springer, 2004.

\bibitem{RimoldiUrbanke:RateSplitting}
B.~Rimoldi and R.~L. Urbanke, ``A rate splitting approach to the {G}aussian
  multiple access channel,'' \emph{IEEE Trans.\ Inf.\ Theory}, vol.~42, pp.
  364--375, 1996.

\bibitem{katz_msc}
G.~Katz, ``On layered transmission in clustered cooperative cellular
  architectures,'' Master's thesis, The Technion---Israel Institute of
  Technology, Aug. 2013.

\end{thebibliography}
\end{document}